# Spin Gap in Optimally-doped YBCO

D. Reznik*[1,2], L. Pintschovius[1], Y. Endoh[3,4], P. Bourges[2], Y. Sidis[2], T. Masui[5], and S. Tajima[5].

[1] Forschungszentrum Karlsruhe, Institut für Festkörperphysik, Postfach 3640, D-76021 Karlsruhe, Germany.
[2] Laboratoire L_eon Brillouin, CEA-CNRS, CE-Saclay, 91191 Gif sur Yvette, France.
[3] Institute for Material Research, Tohoku University, Katahira, Aoba-ku, Sendai, 980-8577, Japan.
[4] International Institute for Advanced Studies, Kizugawadai, Kyoto, 619-0225, Japan.
[5] Superconductivity Research Laboratory, ISTEC, Shinonome, Koutu-ku, Tokyo, 135-0062, Japan.

**Spin gap in optimally-doped YBCO was investigated by inelastic neutron scattering. We found that q-integrated spectral weight in the vicinity of the spin gap has a shape of a step function at 10K with the sharp resolution-limited onset at 33 meV, i.e. there is no observable spectral weight at 31 meV but full spectral weight at 35 meV. The spin gap does not soften or "fill in" as the temperature is increased to $T_c$. Above $T_c$ we find no evidence for q-dependent magnetic signal at 35 meV.**

The origin of an experimentally observed low-temperature energy gap in the dynamical spin susceptibility (spin gap) of superconducting copper oxides is still poorly understood. It scales with $T_c$ reaching just above 30meV in samples with the highest $T_c$. [1,2] We performed detailed measurements of the spin gap in optimally-doped YBCO in light of the recent findings that the magnetic scattering is incommensurate at energies right above the spin gap.

The experiments were performed on the 1T triple-axis spectrometer using the same sample of $YBa_2Cu_3O_{6.95}$ ($T_c$=93K) mounted in the same orientation as in Ref. 3.

Separating the magnetic from the nuclear scattering is the usual challenge of such experiments. We take advantage of the fact that magnetic scattering generally decreases with increasing temperature as the correlations become weaker while the nuclear scattering either stays the same or increases with increasing temperature. In YBCO experience shows that nuclear scattering typically follows the Bose occupation factor as a function of temperature.

To identify magnetic signal above the spin gap we performed scans in **q**-space at 35 meV through the antiferromagnetic wavevector q=(1.5, 0.5, 1.7) at 10K, 100K, and 300K using Cu220 monochromator and PG002 analyzer reflections. The data were divided by the Bose factor to correct for the temperature dependence of one-phonon scattering, which also removed most temperature dependence from the background. An additional small linear term (<5%) was added to the 100K and 300K data in order to match the background on both sides of the scans (near (1.5 0 1.7) and (1.5 1 1.7)). The result is shown in Fig. 1.

The broad asymmetric feature at 100K and 300K has been previously identified as phonon scattering. After the Bose factor correction, it decreases slightly from 100K to 300K as shown in the figure. This small decrease is proportional to the intensity of the phonon feature and could be consistent with the Debye-Waller factor. Thus all **q**-dependent intensity observed at 100K can be accounted for by nuclear scattering, though we cannot rule out a **q**-independent magnetic component. This result is in contrast with underdoped (even $T_c$=89K) YBCO, where significant **q**-dependent magnetic susceptibility exists above $T_c$. Since **q**-dependent magnetic signal disappears almost entirely as T is raised above 100K in our sample, we obtain magnetic intensity at T<$T_c$ by taking the 100K intensity (I(100K)) as background.

Based on this analysis, there are two incommensurate magnetic peaks at 10K that appear on top of the phonon background given by the 100K data. (Fig. 1) It was shown in Ref. 3 that this 10K magnetic scattering between 34 meV and 41 meV originates from a branch of approximately circular symmetry and downward quadratic dispersion from the reduced in-plane wavevector **q**=(0.5, 0.5, 1.7) and energy of ω=41meV. This branch has an abrupt lower cutoff at the spin gap of around 33 meV, which we investigated in detail. The results are presented below.

To study the spin gap we utilized PG002 reflection for the monochromator and analyzer with the final energy fixed at 14.7 meV. Figure 2 shows 10K magnetic intensity, i.e. I(10K) - (I(100K)), at 35 meV compared with the magnetic intensity at 31 meV at 10K and 70K. While there is a clear signal at 35 meV (lineshape is different from Fig. 1 due to a different resolution function), there is no evidence for any appreciable magnetic intensity at 31 meV at either 10K or 70K. Absence of any observable magnetic scattering at 31meV at 70K, the highest temperature where magnetic signal appears above the spin gap [3], shows that the spin gap does not decrease much in energy or "fill up" as the temperature is raised.

Magnetic scattering above the spin gap is incommensurate, so we measured its **Q**-dependence at three incommensurate wavevectors. (Fig. 3) The sharp drop of magnetic intensity occurs at 33 meV at all **Q**. Considering the resolution of 3meV FWHM and taking into account statistical error, we can place the upper bound of 2meV on the intrinsic energy width of the spin gap though it may be even narrower.

To summarize, we found that there is no significant **q**-dependent magnetic intensity above $T_c$ at 35 meV in optimally-doped YBCO. The spin gap appears at 33 meV, is very sharp in energy, **q**-independent, and does not decrease on approaching $T_c$ from below. Another important observation discussed in detail in Ref. [3] is that **q**-integrated magnetic intensity does not decrease on approach to the spin gap, but rather stays constant between 34 and 41 meV. These observations place stringent constraints on possible theories of magnetic fluctuations in the high $T_c$ cuprates.

The crystal growth was supported by the New Energy and Industrial Technology Development Organization (NEDO) as Collaborative Research and Development of Fundamental Technologies for Superconductivity Applications.

*To whom correspondence should be addressed:
E-mail: reznik@llb.saclay.cea.fr

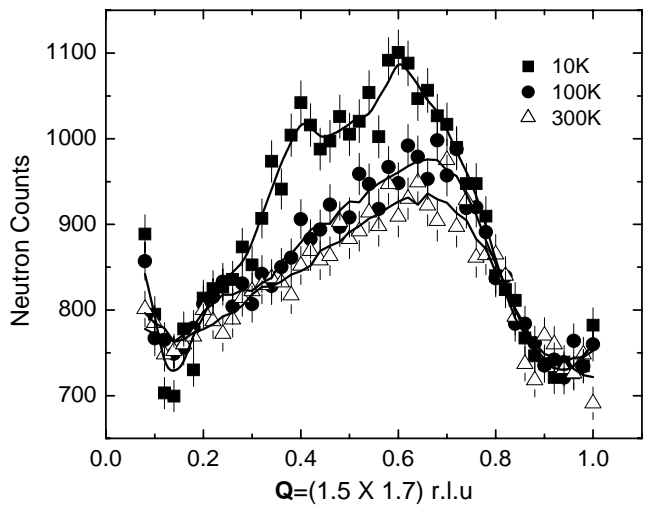

Figure 1: Scans at 35 meV measured with Cu220/PG002 monochromator/analyser and $E_f=30.5$ meV at 10K, 100K, and 300K corrected for the Bose occupation factor. An additional small linear term (<5%) was added to the 100K and 300K data in order to match the intensity on both sides of the scan. Lines are a result of smoothing.

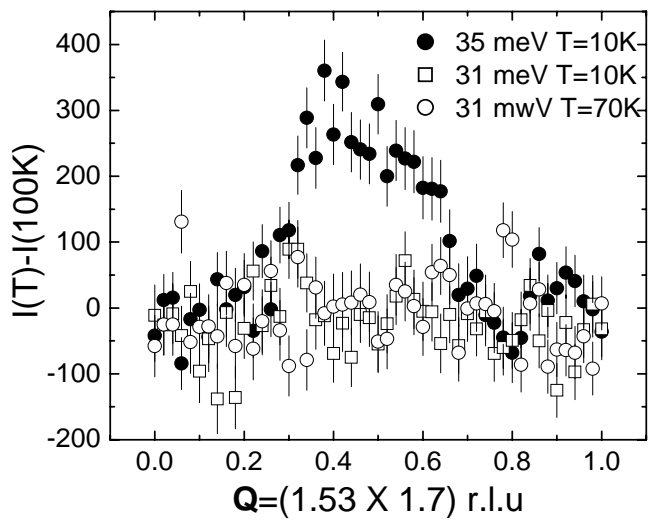

Figure 2: Magnetic neutron scattering at 35 meV and 31 meV. Whereas strong signal is evident at 35 meV, the data at 31 meV is statistically indistinguishable from zero at both 70K and 10K.

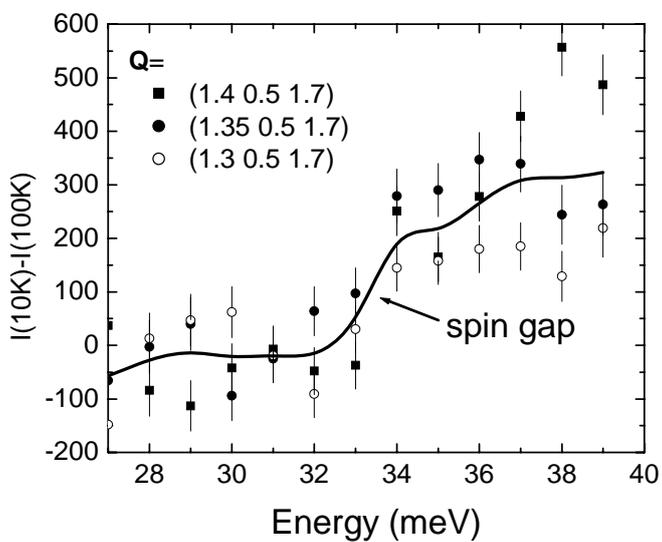

Figure 3: Magnetic scattering cross section as a function of energy at three incommensurate wave vectors. The solid line is a guide to the eye. Spin gap at 33 meV is clearly visible.